\documentclass[twocolumn,11pt]{article}

\usepackage{graphicx}
\usepackage[square]{natbib}
\usepackage{journals1}

\begin{document}

\title{Application of the Allan Variance to Time Series Analysis in Astrometry and Geodesy: A Review}
\author{Zinovy Malkin$^{1,2,3}$\\
$^1$Pulkovo Observatory, St. Petersburg 196140, Russia \\
$^2$St. Petersburg State University, St. Petersburg 199034, Russia\\
$^3$Kazan Federal University, Kazan 420000, Russia\\
e-mail: malkin@gao.spb.ru}
\date{~}
\maketitle

\begin{abstract}
The Allan variance (AVAR) was introduced 50 years ago as a statistical tool for assessing of the frequency standards deviations.
For the past decades, AVAR has increasingly being used in geodesy and astrometry to assess the noise characteristics in geodetic and astrometric
time series.
A specific feature of astrometric and geodetic measurements, as compared with the clock measurements, is that they are generally associated with
uncertainties; thus, an appropriate weighting should be applied during data analysis.
Besides, some physically connected scalar time series naturally form series of multi-dimensional vectors.
For example, three station coordinates time series $X$, $Y$, and $Z$ can be combined to analyze 3D station position variations.
The classical AVAR is not intended for processing unevenly weighted and/or multi-dimensional data.
Therefore, AVAR modifications, namely weighted AVAR (WAVAR), multi-dimensional AVAR (MAVAR), and
weighted multi-dimensional AVAR (WMAVAR), were introduced to overcome these deficiencies.
In this paper, a brief review is given of the experience of using AVAR and its modifications in processing astro-geodetic time series.
\end{abstract}



\section{Introduction}

The noise assessment in physical measurements time series is an important measure of its statistical characteristics and overall quality.
Among the most effective approaches to analyzing measurement noise (scatter) is Allan variance (AVAR), which was originally introduced
to estimate the frequency standards instability \cite{Allan1966}.
Later, AVAR has proved to be a powerful statistical tool for time series analysis, particularly, for the analysis of geodetic and astronomical
observations.
AVAR has been used for quality assessment and improvement of the celestial reference frame (CRF)
\cite{Feissel2000a,Gontier2001,Feissel2003a,Sokolova2007,Malkin2008j,LeBail2010a,LeBail2014a,Malkin2013b,Malkin2015b},
the time series analysis of station position and baseline length
\cite{Malkin2001n,Roberts2002,LeBail2006,Feissel2007,LeBail2007,Gorshkov2012b,Malkin2013b,Khelifa2014},
and studies on the Earth rotation and geodynamics
\cite{Feissel1980,Gambis2002,Feissel2006a,LeBail2012,Malkin2013b,Bizouard2014a}.

AVAR estimates of noise characteristics have important advantages over classical variance estimates such as standard deviation (STD)
and weighted root-mean-square (WRMS) residual.
The latter cannot distinguish the different significant types of noise, which is important in several astro-geodetic tasks.
Another advantage of AVAR is that it is practically independent of the long-term systematic components in the investigated time series.
AVAR can also be used to investigate the spectral characteristics of the signal \cite{Allan1981,Allan1987}
that is actively used for analysis of astrometric and geodetic data \cite{Feissel2000a,Feissel2003a,Feissel2006a,Feissel2007}.

However, the application of original AVAR to the time series analysis of astro-geodetic measurements may not yield satisfactory results.
Unlike clock comparison, geodetic and astrometric measurements mostly consist of data points with unequal uncertainties.
This requires a proper weighting of the measurements during the data analysis.
Moreover, one often deals with multi-dimensional quantities in geodesy and astronomy.
For example, the station coordinates $X$, $Y$, and $Z$ form 3D vector of a geocentric station position (although this example is more complicated
because the vertical and horizontal station displacements caused by the geophysical reasons  may have different statistical characteristics,
including AVAR estimates, see \cite{Malkin2001n,Malkin2013c} and references therein).
The coordinates of a celestial object, right ascension and declination, also form a 2D position vector.
To analyze such data types, AVAR modifications were proposed in \cite{Malkin2008j}, including weighted AVAR (WAVAR), multi-dimensional AVAR (MAVAR),
and weighted multi-dimensional AVAR (WMAVAR).
These modifications should be distinguished from the classical modified AVAR introduced in \cite{Allan1981}.

The rest of the paper is organized as follows.
Section~\ref{sect:overview} introduces AVAR and its modification, and gives several practical illustrations of their basic features.
In Section~\ref{sect:results}, a brief overview is provided of the works that employ AVAR in geodesy and astrometry, and
basic results obtained with the AVAR technique are presented.
Additional details and discussion on the use of AVAR in space geodesy and astrometry can be found in \cite{LeBail2004t,Malkin2011c,Malkin2013b}.


\section{Overview of AVAR and its modifications}
\label{sect:overview}

The classical time-domain AVAR applied to the time series $y_i, i=1, \dots, n$ is given by \cite{Allan1966}
\begin{equation}
AVAR = \frac{1}{2(n-1)}\sum_{i=1}^{n-1}(y_i-y_{i+1})^2\,.
\label{eq:AVAR}
\end{equation}

Allan deviation ADEV = $\sqrt{\rm{AVAR}}$ is used as a noise characteristics in many data analysis applications.
Both AVAR and ADEV estimates will be used throughout the paper depending on the context.

In metrology, the analyst is normally interested in computing not only the parameter under investigation but also its uncertainty,
as a measure of reliability of obtained result.
A method of estimating the AVAR confidence interval is proposed in \cite{Howe1981}.
The method of estimating the AVAR confidence interval proposed in \cite{Howe1981} was applied in \cite{LeBail2006} for analysis of geodetic time series.

The original AVAR definition supposes that all measurements (observations) have uniform precision.
In most geodetic and astrometric applications, however, measurements display different level of precision.
Therefore, appropriate weighting is necessary during analysis.
Given the measurements $y_1,y_2,\dots,y_n$ with the associated uncertainties $s_1,s_2,\dots,s_n$
Then the WAVAR estimate can be defined to treat unequally weighted data \cite{Malkin2008j}:
\begin{equation}
\begin{array}{rl}
& WAVAR = \frac{\displaystyle 1}{\displaystyle 2p}
   \displaystyle\sum_{i=1}^{n-1}p_i(y_i-y_{i+1})^2 \,, \\[2ex]
& p = \displaystyle\sum_{i=1}^{n-1}{p_i}\,, \\[4ex]
& p_i = (s_i^2+s_{i+1}^2)^{-1} \,,
\end{array}
\label{eq:WAVAR}
\end{equation}
where $p_i$ are the weights.

\begin{figure}[ht!]
\centering
\includegraphics[width=\columnwidth]{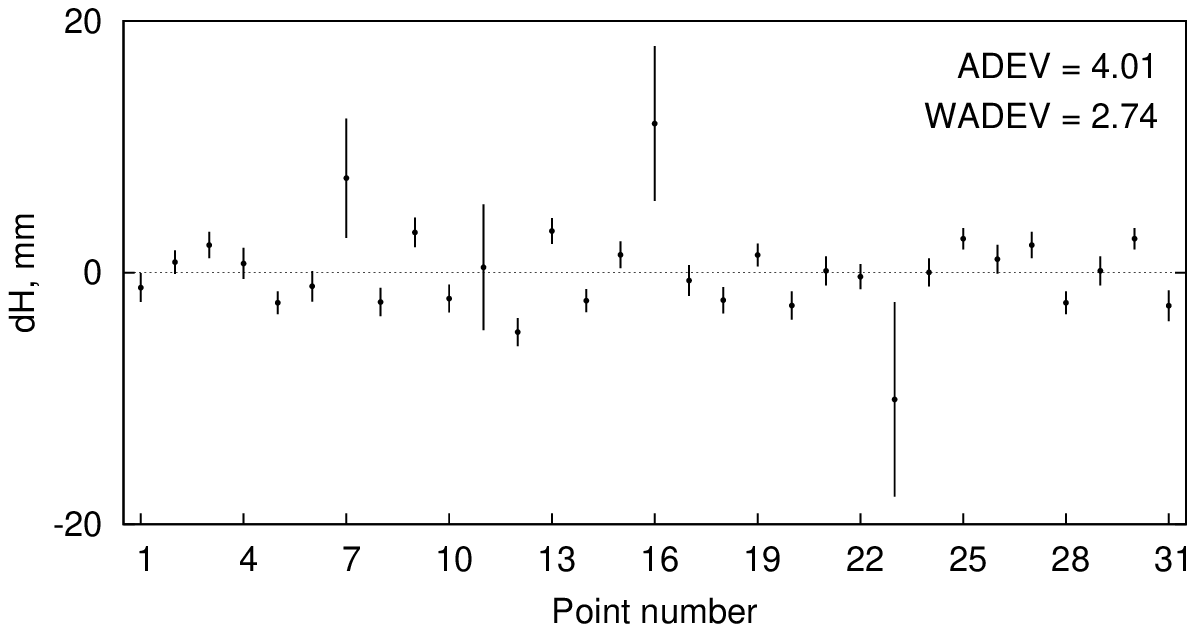}
\includegraphics[width=\columnwidth]{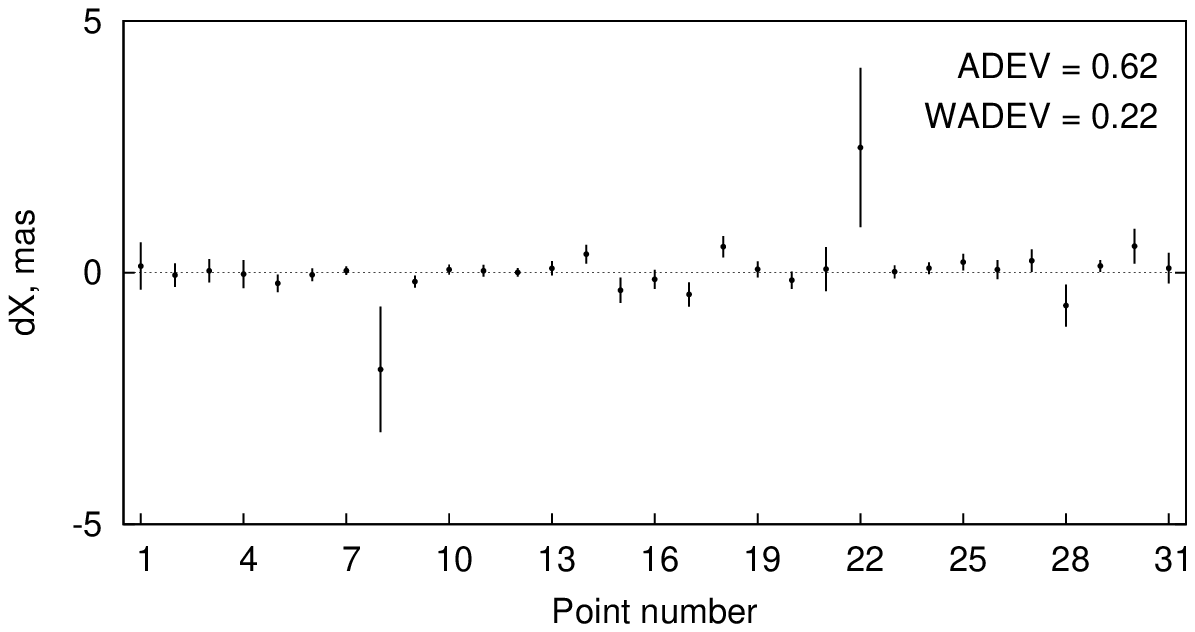}
\includegraphics[width=\columnwidth]{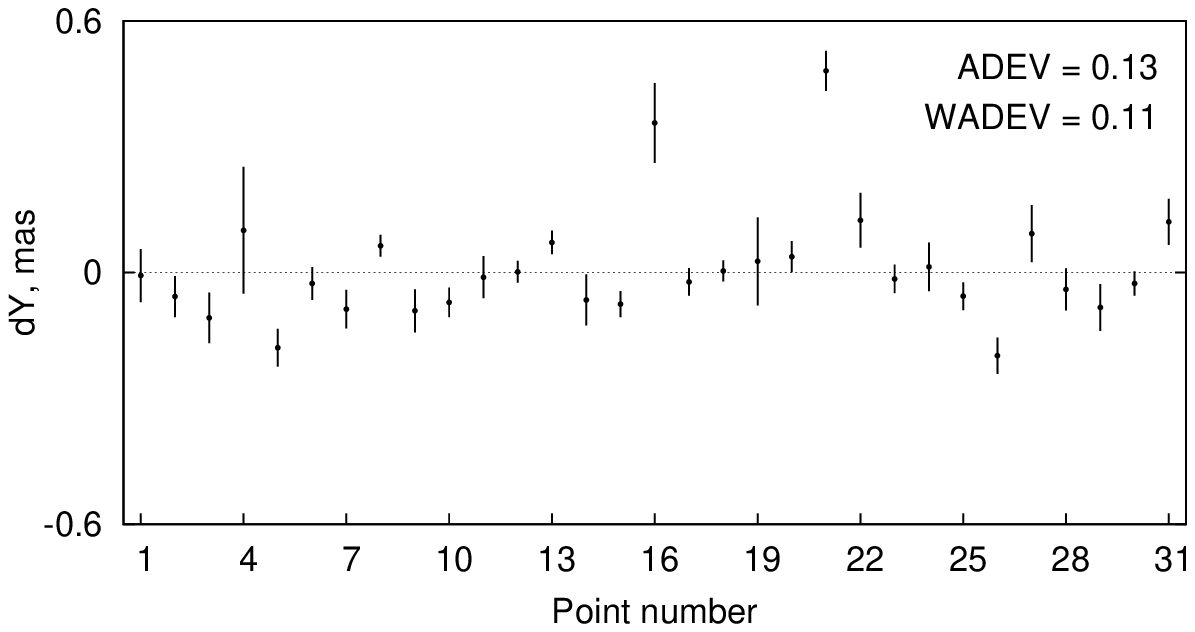}
\caption{Three examples of time series of unequally weighted measurements: station height (top) and celestial pole offset (middle and bottom).
Two upper cases show series that contain outliers with large measurement errors; for these series the outliers substantially impact ADEV, and
WADEV estimate looks more realistic.
Third series contains two outliers with uncertainties similar to that of other measurements; in this case WADEV estimate has no advantage over ADEV.}
\label{fig:WAVAR_example}
\end{figure}

Figure~\ref{fig:WAVAR_example} illustrates the difference between AVAR and WAVAR for three time series with outliers:
daily estimates of the station height and two series of celestial pole offsets (CPO).
The latter are the differences in $dX$ and $dY$ between the observed $X$ and $Y$ coordinates of the celestial pole from the celestial pole position
defined by the International Astronomical Union IAU 2000/2006 precession-nutation model \citep{IERSConv2010} as measured by 
VLBI (very long baseline interferometry).
Here, we consider points with abnormally large deviations from the mean and/or with abnormally large uncertainties to be outliers.
Analysts usually aim to identify and reject outliers before conducting statistical analysis; however, generally this is not a simple problem.
One can see from Fig.~\ref{fig:WAVAR_example} that WADEV can provide more robust estimate even in the presence of outliers  that were not previously
detected and eliminated.
In other words, WADEV is less dependent on the quality of the outlier detection procedure used prior to data analysis.
Notably, WADEV dependence on outliers is low when the outliers exhibit greater uncertainties than the uncertainties of ``good'' measurements.
This is mostly the case in practice.
Exceptions are possible, as shown in the bottom plot of Fig.~\ref{fig:WAVAR_example} although they are rare.

\begin{figure}[ht!]
\centering
\includegraphics[width=\columnwidth]{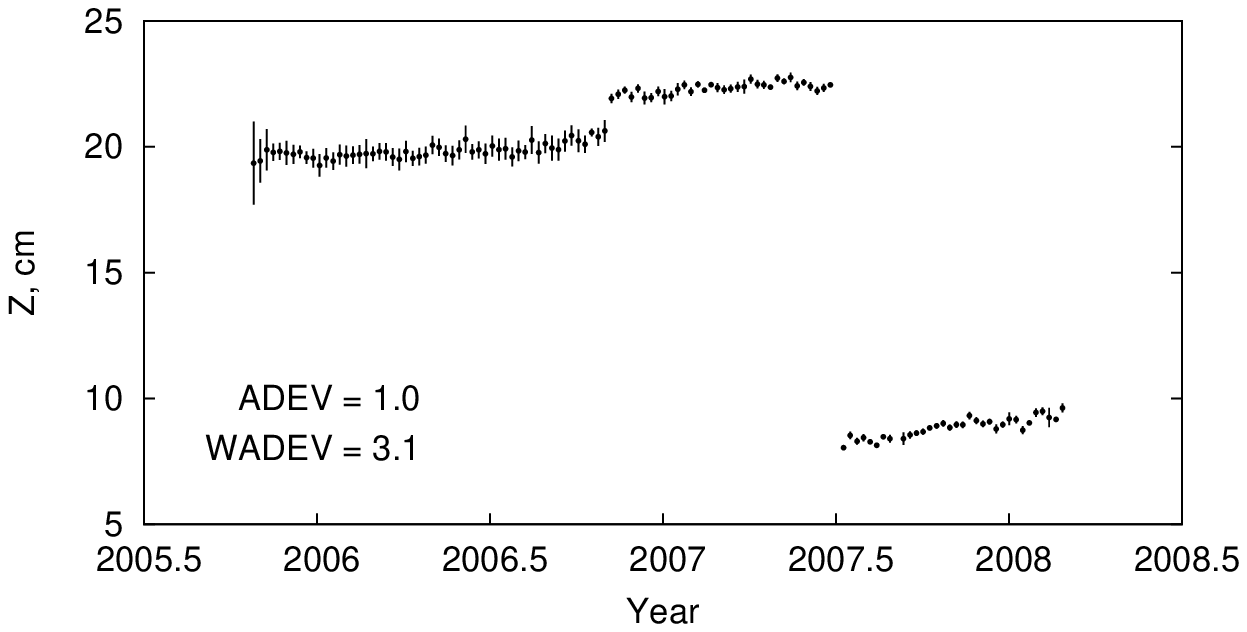}
\includegraphics[width=\columnwidth]{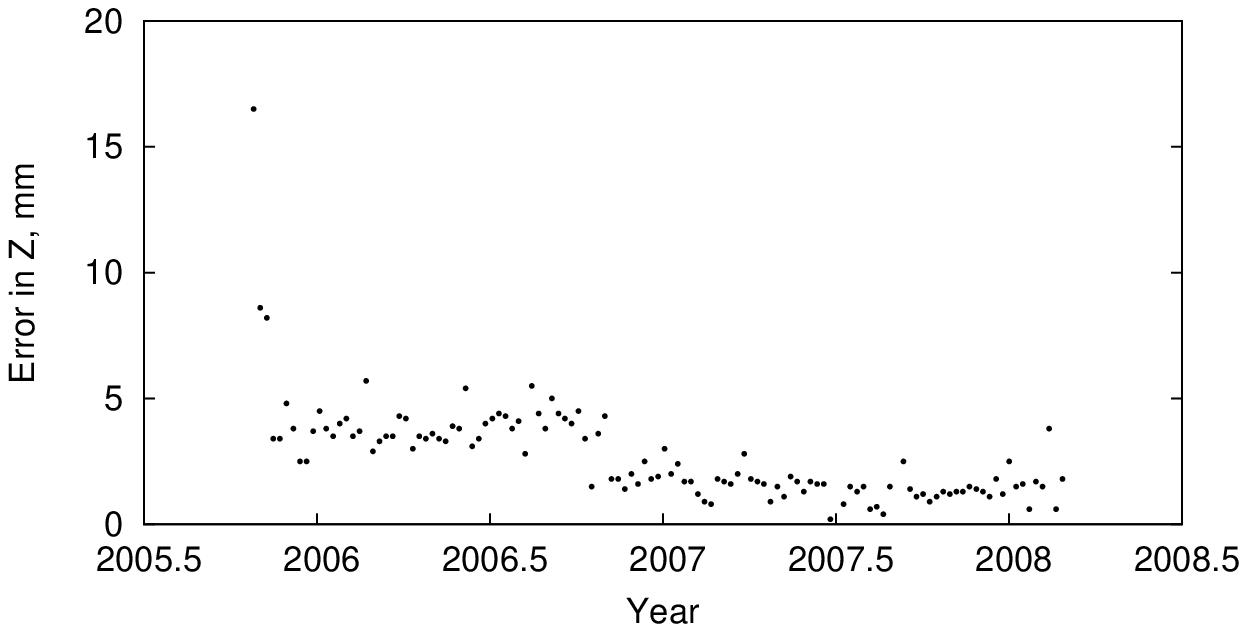}
\caption{Time variation of Z coordinate of a GPS station and its uncertainty.}
\label{fig:hflk}
\end{figure}

It is important to keep in mind that AVAR, as well as any other statistics has limitations depending on data under analysis.
Let us consider an example of time series of Z coordinate of the station position presented in Fig.~\ref{fig:hflk}, upper plot.
One can see that both ADEV and WADEV give unsatisfactory results.
Both estimates are affected by jumps in the time series and are too large as compared with the actual measurement noise.
Moreover, WADEV gave even worse result than ADEV.
The reason of such a confusing WADEV estimate is the following.
As can be clearly seen in Fig.~\ref{fig:hflk}, bottom plot, the measurement errors changed with time.
In particular, the smallest errors are observed in the second part of the time series, where the largest jump occurred.
As a result, the measurements made around the second, largest jump were entered to the WADEV computation with larger weight.
For comparison, if only the part of the time series after the last jump (epoch 2007.5) is used for analysis, ADEV = WADEV = 0.2~cm.

So, ADEV, as well as most of other statistics, cannot provide a satisfactory result for non-stationary time series.
In particular, if the time series under investigation has heterogeneous uncertainty, the use of dynamic AVAR (DAVAR) \cite{Galleani2009}
is clearly advantageous.
The DAVAR is defined as AVAR computed over a sliding window moving along the time series, and thus it allows estimating the variation of the
time series noise characteristics with time. 

In astro-geodetic time series analysis, we often address multi-dimensional values, such as geocentric station coordinates $X$, $Y$, and $Z$ (3D vector)
and the position of celestial objects (right ascension and declination) in the sky (2D unit vector).
Further extension of the WAVAR estimator WMAVAR can facilitate the analysis of such data \cite{Malkin2008j}.
Given $k$-dimensional vector of measurements $y_i=(y_i^1,y_i^2,\dots,y_i^k)$ 
with the $k$-dimensional vector of corresponding uncertainties $s_i=(s_i^1,s_i^2,\dots,s_i^k)$,
WMAVAR considers the measurements to be the points in the $k$-dimensional space.
Then, the difference between adjacent measurements is taken as the Euclidean length between these values.
Thus, the $k$-dimensional WMAVAR estimate can be defined as
\begin{equation}
\begin{array}{rl}
& WMAVAR = \frac{\displaystyle 1}{\displaystyle 2p}
  \displaystyle\sum_{i=1}^{n-1}p_i \, d_i^2\,, \\
& d_i = |y_i-y_{i+1}|\,, \\[1ex]
& p = \displaystyle\sum_{i=1}^{n-1}{p_i}\,, \\[3ex]
& p_i = \left( \displaystyle\sum\limits_{j=1}^k
  \left[(y_i^j-y_{i+1}^j)/d_i\right]^2 \left[(s_i^j)^2+(s_{i+1}^j)^2\right] \right)^{-1} \,,
\end{array}
\label{eq:WMAVAR}
\end{equation}
where $|\dots|$ denotes the Euclidean norm, and $p_i$ are the weights.
The expression for $p_i$ is derived from classical error propagation low and is therefore theoretically correct.
However, this expression cannot be applied in practice because it generates a singular when $d_i=0$, that is, a case in which adjacent measurements
are equal.
This situation is not extraordinary.
To eliminate this problem, a simplified formula was proposed in \cite{Malkin2008j}:
\begin{equation}
p_i=\left( \displaystyle\sum\limits_{j=1}^k \left[(s_i^j)^2 + (s_{i+1}^j)^2 \right] \right)^{-1} \,.
\label{eq:p_alt}
\end{equation}
Numerous tests with real data revealed no practical difference between the use of Eqs.~(\ref{eq:WMAVAR}) and that of (\ref{eq:p_alt})
in computing for WMAVAR.

It is easy to see that WMAVAR (WMADEV) is a universal definition that encompasses all other variants such as AVAR (ADEV), WAVAR (WADEV), and
MAVAR (MADEV) as special cases.
It can be also noted that for a time series whose $s_i$ values are close to one another WMADEV is approximately 
$\sqrt{k}$ times as large as the WADEV computed for one (each) vector dimension.
For example, given three series of equally precise measurements of station coordinates $X$, $Y$, and $Z$ having about the same WADEV estimate
$\sigma_1$, the 3D WMADEV estimate will be equal to roughly $\sqrt{3}\sigma_1$.

\begin{figure*}[ht!]
\centering
\includegraphics[width=\hsize]{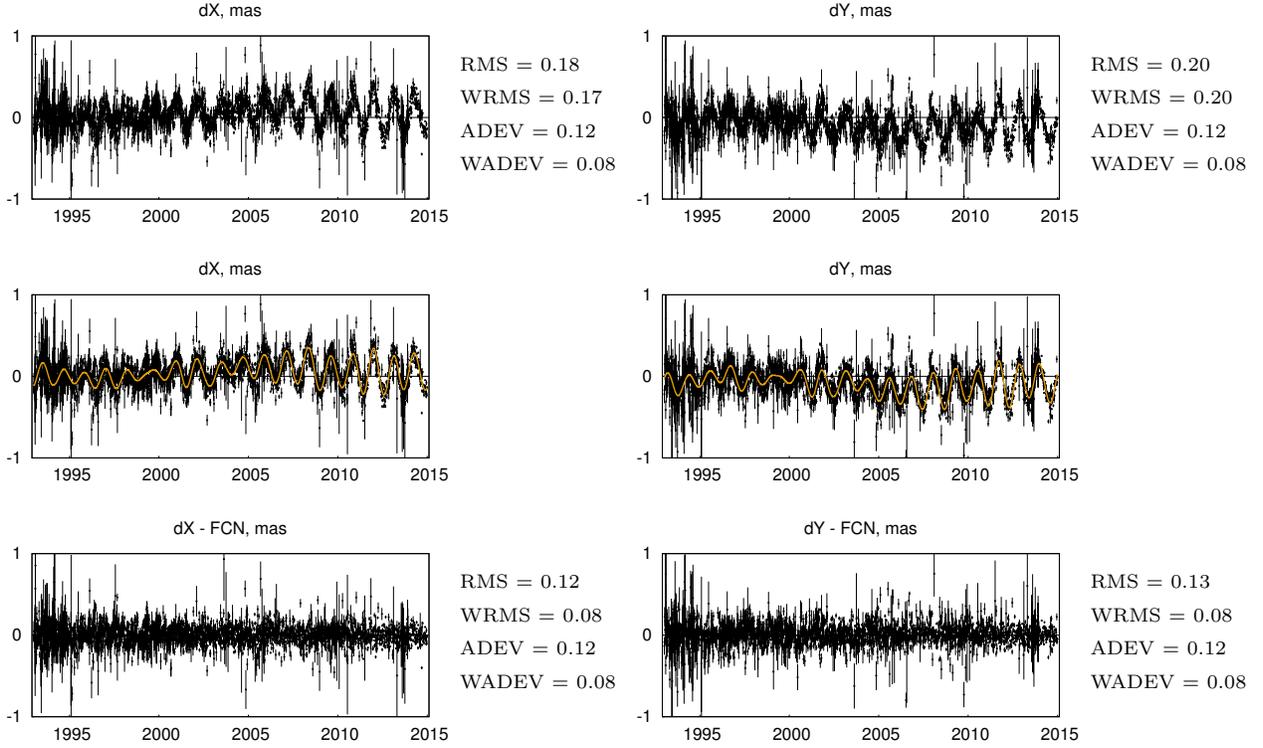}
\caption{The original IVS CPO series (upper row), the same series with the ZM4 CPO model (middle row), and the residuals between CPO and ZM4 model
(lower row). Different scatter statistics are presented to the right of each series.}
\label{fig:cpo_fcn}
\end{figure*}

A main advantage of AVAR when used as a noise level estimator is its weak dependence on low-frequency variations.
Figure~\ref{fig:cpo_fcn} depicts the results of the CPO determination in the International VLBI Service for Geodesy and astrometry (IVS) \cite{Schuh2012,Boeckmann2010}.
One can see that the CPO series include quasi-harmonic signal with period about 430 days corresponding to the free Earth's core nutation (FCN) and
low-frequency components caused by the errors in the precession-nutation model.
In this study, CPO is modeled by the ZM2 and ZM4 models \cite{Malkin2007i,Malkin2013d}
The results presented in Fig.~\ref{fig:cpo_fcn} show that the WRMS estimate depends heavily on the low-frequency components of the analyzed signal.
The removal of these components requires highly accurate modeling of the underlying physical processes.
By contrast, AVAR is practically unaffected by the presence of trend and low-frequency harmonics.

An important AVAR application is in characterizing the spectral behavior of a time series \cite{Allan1981,Allan1987}.
The assessment of noise spectral characteristics may be very important in astrometric and geodetic data analyses.
For example, a choice of the method for computing the station velocity uncertainty from the station position time series crucially depends
on the noise spectral type \cite{Mao1999,Williams2003}.
Knowledge of the spectral noise structure facilitates the construction of a proper covariance matrix of a geodetic signal \cite{Zhang1997}.

As with clock deviations, the noise processes in geodetic and astronomical time series can generally be described by a power law
\begin{equation}
S_y(f) = S_0 \, f^\alpha \,,
\label{eq:power_law}
\end{equation}
where $S_0$ is a normalizing constant and $\alpha$ is the spectral index.

It was supposed above that $\{y_i\}$ are the original measurements and that AVAR is just used as a scatter level estimate.
The original AVAR definition as given in full by \cite{Allan1966} is
\begin{equation}
AVAR = \sigma^2(\tau) = \frac{1}{2} \langle (\bar y_i - \bar y_{i+1})^2 \rangle \,,
\label{eq:original_AVAR}
\end{equation}
where $\bar y_i$ are averaged values over the sampling interval (number of points) $\tau$.
Then, the following relation can be used to identify noise spectral characteristics:
\begin{equation}
\log(\sigma^2(\tau)) = \mu \, \log(\tau) \,.
\label{eq:avar_tau}
\end{equation}

\begin{figure}[ht!]
\centering
\includegraphics[width=0.9\columnwidth]{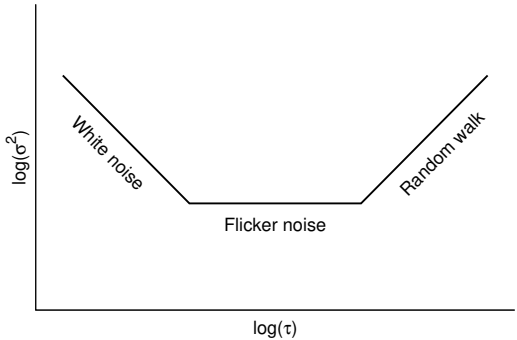}
\caption{Dependence of AVAR on the sampling interval $\tau$ for the three most common typical noise types.}
\label{fig:noise_type}
\end{figure}

The $\mu$ value is connected to the spectral index by $\alpha=-(\mu+1)$ for $-3 < \alpha < 1$ ($-2 < \mu < 2$) \cite{Allan1981}.
Then, the prevailing noise type in the time series can by classified as white noise ($\mu=-1, \alpha=0$), flicker noise ($\mu=0, \alpha=-1$),
random walk ($\mu=1, \alpha=-2$), or another noise type (see Fig.~\ref{fig:noise_type} for illustration).
It must be kept in mind, however, that the analogy of the AVAR-derived log-log slope $\mu$ with the spectral index is only justified for
stationary stochastic processes.

Given real signals, noise often exhibits different spectral characteristics at various frequency bands; thus, a full noise spectrum consists
of several components: 
\begin{equation}
S_y = \sum_{j=1}^{N_f}{c_j\,f_j^{\alpha_j} \,.}
\label{eq:complex_spectra}
\end{equation}

It must be noted that this noise classification method must be practiced with care. 
First, the analyzed time series should be sufficiently long; at least several hundreds of measurements are needed to obtain a reliable result
\cite{Williams2003}.
Moreover, trend, seasonal and other long-periodic terms (beyond the frequency band of the analyst's interest) from the analyzed signal should be
removed prior to spectral analysis; otherwise, the log-log spectrum (\ref{eq:avar_tau}) can be distorted \cite{LeBail2006,LeBail2012,Khelifa2014}.

In case of uniform noise, computing and analyzing the classical power spectra may be preferable, such as through Fourier transform.
In case of mixed noise, however, AVAR analysis can provide valuable supplemental information for adjusting the analysis procedure. 

Indeed, the use of the AVAR algorithm in characterizing the spectral behavior of a time series can easily be generalized for weighted and
multi-dimensional measurements through the application of WMAVAR and weighted average at each sampling interval $\tau$.
It can be mentioned here that the computation of weighted average uncertainty does not generate an unambiguous solution.
A possible approach to solving this problem in practical applications is discussed in \cite{Malkin2013e}.
  
Among other applications, AVAR can be used to estimate the Hurst parameter \cite{Bregni2008}.
The Hurst parameter (or Hurst exponent) $H$ is related to such time series statistical properties as
long-term memory, self-similarity (fractal dimension) and spectral index.

Notably, the original AVAR is defined for evenly spaced time series, which generally does not apply to all geodetic and especially astronomical data.
The simplest method of solving this problem is to form so-called normal points that average data over equally spaced intervals.
Such a method was used in \cite{Feissel2003a} for radio source position time series and in \cite{Feissel2007} for station position time series.
Nonetheless, the process of averaging the original series inevitably leads to loss of information regarding high-frequency noise characteristics.
However, if AVAR is merely used as a measure of random deviation level, uneven data spacing should not influence analysis results, supposing
stationary noise.
A special type of unevenly spaced data is the time series that consists of equally spaced measurements with gaps.
In \cite{LeBail2006}, a special technique has been developed to fill in data gaps prior to statistical analysis.
Additional considerations for the application of AVAR to series with gaps are given in \cite{Sesia2008}.


\section{Use of AVAR in geodesy and astrometry}
\label{sect:results}

This section provides an overview of the applications of AVAR in astrometry, geodesy, and geodynamics.
In these fields, AVAR is mainly applied to investigate space geodesy station displacements and the stability of VLBI-derived celestial object positions.
Several examples of AVAR applications are discussed in the following subsections to effectively illustrate some features of AVAR estimates. 

\subsection{Celestial reference frame}

The studies described in this subsection are CRF-related investigations, such as the analysis of radio source coordinates
catalogs and the improvement of the International Celestial Reference Frame (ICRF) through the detection of radio sources with unstable positional
behavior. 

Accuracy assessment of the CRF catalogs is not a trivial task because only the differences between the radio source positions in various catalogs
are known (this situation is very similar to the case of clock comparison).
A possible absolute method of catalog comparison is based on the comparison of the noise levels in the CPO series as computed with the compared
source catalogs \cite{Malkin2008j}.
In this study, the radio source position catalogues obtained in 8 IVS analysis centers were compared by means of two scatter indices.
Both indices are based on the analysis of the CPO series computed with different catalogs.
The first index is computed from analysis of the residuals of CPO series with respect to the IAU 2000/2006 precession-nutation model.
This index occurred to be less sensitive, and, besides, it is not fully independent, since the IAU model was determined form VLBI data analysis
using some CRF realizations under study.
Conversely, WADEV and WMADEV indices provide an independent estimate of the quality of CRF realizations.
The WMADEV occurred to be most sensitive to differences in CRF realizations, and thus can be considered as a preferred scatter index.

With the use of this method, the ICRF catalog \cite{Ma1998,Fey2004a}, hereafter referred to as ICRF1, was compared with the
catalog RSC(PUL)07C02 derived at the Pulkovo observatory \cite{Sokolova2007}.
The latter was computed as combined catalog using several individual catalogs obtained in the IVS analysis centers in 2005.
It uses more observations than ICRF1 catalog computed in 1995, and a new method of determination and elimination of the systematic errors in source
positions.
Two CPO time series for 2002--2006 obtained with two CRF realizations were computed.
The results of comparison are presented in Table~\ref{tab:cpo_cat} and show that the noise level of the CPO series computed
with the Pulkovo catalog is lower than that of the series computed with ICRF1
This finding suggests that the catalog RSC(PUL)07C02 was more accurate than ICRF1.
This conclusion was later confirmed by improvement of the EOP results obtained with the new catalog. 
The analysis of the astrometric aspects of catalog computation and comparison is beyond the scope of this paper.
It is only important here to show that estimation of the noise level in the CPO time series can serve as an external characterization of the
accuracy of radio source catalogs (CRF realizations).

Subsequently, a similar comparison was made in \cite{Malkin2013b} between two ICRF realizations, ICRF1 \cite{Fey2004a}
and ICRF2 \cite{Malkin2015c}.
The latter was computed in 2009 and using about 6.5~mln observations collected during 1979--2009 as compared with 1.6~mln observations
collected in 1979--1995 used for ICRF1.
As a result, several direct catalog comparisons showed much better precision and of ICRF2 with respect to ICRF1 \cite{Malkin2015c}.
Two CPO time series for 2002--2010 obtained with two CRF realizations were computed.
The result of the CPO noise analysis made in \cite{Malkin2013b} and presented in Table~\ref{tab:icrf2-icrf1} independently confirmed that ICRF2
is more accurate than ICRF1.

\begin{table}
\centering
\caption{Noise level in the CPO time series computed with Pulkovo and ICRF1 catalogs of radio source positions as estimated with weighted AVAR.}
\label{tab:cpo_cat}
\begin{tabular}{cccc}
\hline
Catalog & \multicolumn{2}{c}{WADEV} & WMADEV \\
        & $dX$    & $dY$    &  \\
        & $\mu$as & $\mu$as & $\mu$as \\
\hline
ICRF1   & 112.8 & 108.7 & 168.1 \\
Pulkovo & 105.0 & 105.8 & 160.9 \\
\hline
\end{tabular}
\end{table}

\begin{table}
\centering
\caption{Noise level in the CPO time series computed with ICRF1 and ICRF2 catalogs of radio source positions as estimated with weighted AVAR.}
\label{tab:icrf2-icrf1}
\begin{tabular}{cccc}
\hline
Catalog & \multicolumn{2}{c}{WADEV} & WMADEV \\
        & $dX$    & $dY$    &  \\
        & $\mu$as & $\mu$as & $\mu$as \\
\hline
ICRF1 & 101.6 & 107.3 & 148.4 \\
ICRF2 & ~91.7 & ~89.2 & 128.3 \\
\hline
\end{tabular}
\end{table}

It should be realized that both $dX$ and $dY$ are merely the spherical coordinates of the celestial pole, and their statistical behaviors
are closely correlated.
Thus, the 2D WMADEV estimate serves as a compact and convenient tool for describing the noise component in CPO (and other similar) series. 

The noise characteristics of a source position time series can be used for source ranking in terms of temporal and spatial stabilities,
as well as for compiling a list of sources that are not stable enough to be included in VLBI global solution and thus requires special handling.
The position instabilities of 16 radio sources with long and dense observational history were investigated in detail,
including the spectral characteristics of the source position time series through the AVAR analysis \cite{Feissel2000a}.
The analysis results showed that the time variabilities of these objects exhibit varied spectral characteristics that may be ascribed
to diversified physical processes. 

Several statistical techniques were applied to source position time series to assess the VLBI-derived CRF in \cite{Gontier2001}.
In particular, AVAR analysis was performed to assess the noise spectrum of 60 source position time series.
Sampling interval $\tau$ varied from the initial time span of the series to 1/4 $\ldots$ 1/3 of the data span (six months to four years).
It was found in \cite{Gontier2001} that roughly 60\% of the series predominantly exhibited white noise and 40\% of the series
mainly displayed flicker noise.
Approximately 160 sources have sufficiently long and dense observation histories from which the AVAR can be derived for a one-year sampling interval.
This information was employed to investigate source position stability further. 

An extended similar work was conducted to select a set of positionally stable sources through an analysis of source position time series with
the use of WRMS and ADEV \cite{Feissel2003a}. 
This study also included an analysis of the apparent drifts in radio source positions.
In this study, 707 sources were analyzed, and 199 were recommended as candidate core sources for the subsequent ICRF realization.
In particular, roughly 60\%, 35\%, and 5\% of the series exhibited white noise, flicker noise, and random work noise, respectively. 

The methods developed in \cite{Feissel2003a} for source stability characterization were also used to analyze recent source position time series
\cite{LeBail2010a}.
This study investigated the position stability of ICRF2 sources using AVAR and considered the apparent linear source motions.
It was found that the selection of stable core sources depends on the configuration of the VLBI solution used to derive the position time series
and may differ by up to 20\% of the selected sources. 
Furthermore, source stability found in \cite{LeBail2010a} improved significantly with respect to that found in \cite{Feissel2003a}
because of the advances in observing technology as well as the increase in observing networks, which directly affect the precision and accuracy 
of VLBI results \cite{Malkin2009g}.
Another analysis performed in \cite{LeBail2010a} found that the log-log slope $\mu$ computed according Eq.~(\ref{eq:avar_tau}) may not remain constant
with time for long radio source position time series.
Subsequently, this work was extended by incorporating the most recent observations in the analysis \cite{LeBail2014a}.

AVAR was also applied to the time series analysis of 15 source positions provided by 9 IVS analysis centers in the framework of ICRF2 preparation
campaign \cite{Malkin2015b}.
In particular, it was found that AVAR generates an inadequate estimate of noise level when jumps are present in a time series, as discussed above.
In general cases, a composite index of WRMS and WADEV can be applied to enhance the robustness of the measure of source position stability,
as previously discussed in \cite{Feissel2003a}.

\subsection{Geodesy}

Many studies have confirmed that AVAR is an effective tool for investigating the noise characteristics of station position time series,
particularly those obtained with space geodetic methods, such as VLBI, GPS, satellite laser ranging (SLR), and Doppler orbitography and
radiopositioning integrated by satellite (DORIS). 

AVAR was used to estimate the week-to-week repeatability of the station position of the European Permanent GPS Network (EPN) station position
in the presence of seasonal station movements \cite{Malkin2001n}.
The use of AVAR in this case is reasonable because AVAR does not require the estimation of a systematic trend and seasonal variations in station
positions, which are difficult to model.
In the process, the WRMS noise estimate that was previously used to compare station position time series is distorted.

AVAR was also applied to the time series analysis for baseline length in relation to GPS-based deformation monitoring applications \cite{Roberts2002}, 
where this approach was used to determine the temporal upper limit accuracy of each monitoring system and to help distinguish a systematic error from
a genuine geophysical signal for a new measurement.

The noise spectrum in DORIS ground station motion was investigated by applying AVAR to the decomposition of the 3D signal into its principal
components in the time domain \cite{LeBail2006}.
The noise level was estimated, and white noise was dominant after the linear drift in the series of station coordinates was eliminated.
In addition, the AVAR signature was sensitive to periodic terms.
To avoid bias in the estimation of noise type, three periodic terms, namely annual, semi-annual, and 117.3-day, were filtered out prior to
computation of AVAR.
The annual and semiannual seasonal variations are usually observed in the station position series derived from astronomical and space geodesy
observations see \cite{Dong2002,Malkin2013c} and literature cited therein.
The 117.3-day signal is specific for DORIS observations.
It is connected with the 117.3-day period in orbit node movement for the TOPEX/Poseidon satellite extensively used during many years for DORIS.
Besides, the method of estimating the AVAR confidence interval proposed in \cite{Howe1981} was applied in \cite{LeBail2006}
to the DORIS stations position time series.
This became usual practice in future works \cite{LeBail2010a,LeBail2012,LeBail2014a}. 

The time series of the station coordinates obtained from various space geodetic observations were analyzed in \cite{LeBail2007}.
It was found that the station position series obtained from VLBI, SLR, and DORIS contain white noise, whereas the majority of GPS
station motions exhibit flicker noise.
The atmospheric loading time series (change in the station positions caused by variations of the atmospheric pressure) also contain white noise,
as do the series of the transformation parameters in the midst of position time series obtained by different techniques.
A similar but more detailed analysis was presented in \cite{Feissel2007}.

The noise and low-frequency components of coordinate variations were investigated with the data obtained from three permanent GPS stations
located in the territory of Pulkovo Observatory \cite{Gorshkov2012b}.
AVAR was used in this work to assess the noise in the station position time series.
The results were compared with the spectral index obtained from Hurst exponent analysis.
Both methods suggested that the position time series for the three Pulkovo stations are generally dominated by flicker and mixed white-flicker noises.

An AVAR analysis of the noise characteristics in the time series for DORIS station position was presented in \cite{Khelifa2014}.
Once the trends and seasonal components from the analyzed time series were eliminated, the results of the subsequent AVAR analysis indicated
that the three solutions are dominated by white noise in all three components (north, east, and vertical).
Upon confirming the dominant noise type in the position time series, an adequate method for further analysis was chosen.

\subsection{Earth's rotation and geodynamics}

Earth orientation parameters (EOP) define the rotation of the terrestrial reference system ITRS with respect to the celestial reference system ICRS
\citep{IERSConv2010}.
These parameters include polar motion (movement of the Earth's rotation axis in ITRS), precession-nutation
(movement of the Earth's rotation axis in ICRS), and Universal Time, which depends on the rotational speed of the Earth around its axis.
EOP are determined through the same space geodetic methods described in the previous subsection. 
The final EOP series for practical use are derived by combining the results obtained at different EOP analysis centers.
The noise characteristics of EOP series should be investigated for different tasks, such as quality assessment of the EOP series and the weighting
of these series during combination. 

The computation method of the combined EOP series at the Bureau International de L'Heure (BIH) involved an AVAR analysis of the input EOP
series \cite{Feissel1980}.
Specifically, AVAR was used to obtain the frequency dependence of the noise power spectrum in the combined series.
In particular, it was found that the yearly bias in a time series with respect to the combined series is dominated by flicker noise,
which may indicate accidental local and instrumental changes.
However, the individual results obtained at separate stations are not stable enough; thus, corrections for individual stations cannot
be reliably applied. 

AVAR analysis is also a main step in the EOP combination procedure employed by the International Earth Rotation and Reference Systems Service (IERS),
which succeeded BIH \cite{Gambis2002}.
AVAR is used to characterize the internal random deviations in the EOP time series at various time scales.
White noise and flicker noise dominate these series
Another application of the AVAR is in comparing EOP series to evaluate their internal precision.
These data are used to assign proper weights to EOP series during combination. 

An analysis of several CPO series computed at different IVS analysis centers was performed in \cite{Malkin2013b}.
The results of this study confirmed that the WRMS estimate of the noise in CPO series is heavily dependent on the model of low-frequency
variations, whereas WADEV is practically independent of such a model.
The use of WADEV estimates also mitigated the influence of outliers significantly.
Using WADEV estimates also allowed us to severely mitigate the influence of outliers. 
Another result of the study is that 2D WMADEV estimates are close to the averaged WADEV estimates computed for
$dX$ and $dY$ and multiplied by $\sqrt{2}$, as was discussed in Section~\ref{sect:overview}.

Seasonal and non-seasonal components in the geocenter motion signal measured by satellite geodesy techniques were analyzed to determine
the connection of these components with geophysical fluids \cite{Feissel2006a}.
AVAR was used in this study to compute and compare the spectral characteristics of geodetic and geophysical geocenter time series over
a sampling period of one month to one year.
In particular, the dependence of the log-log $\tau$-AVAR pattern on the seasonal components in the analyzed signal was determined.

Geodetic and modeled excitation functions were compared via AVAR in \cite{Bizouard2014a}.
In this study AVAR analysis was applied to Earth's rotation excitation functions, complementing other methods of statistical analysis.
Bias and seasonal variations were removed from the signals prior to AVAR computation.
Comparison of ADEV obtained for geodetic and geophysical excitations series, and their residuals for different sample interval $\tau$
allowed to draw a conclusion on the geophysical excitation of the Earth's rotation.


\section{Conclusions}

AVAR is an effective statistical tool for analyzing the time series of observational data in astronomy and geodesy, as well as of all other time series.
Important independent characteristics of the noise component in the studied signal can be obtained through this tool. 
The main application of AVAR to time series analysis is in determining the signal scatter level and spectral analysis with the primary aim
to identify the dominant noise type in a time series.
This information can be applied to refined data analysis, such as computing realistic uncertainty of station velocities \cite{Mao1999,Williams2003})
and regularization of EOP series \cite{LeBail2012}.

AVAR modifications, namely, WAVAR, MAVAR, and WMAVAR, were proposed for processing weighted and/or multi-dimensional data \cite{Malkin2008j}.
These modifications serve as effective and convenient tools for data analysis in geodesy and astronomy. 
WMAVAR is the most general estimator and encompasses AVAR, WAVAR, and MAVAR as special cases.
In particular, 2D WMAVAR can be used for complex data processing.
Indeed, the AVAR modifications we proposed should not be confused with the ``original'' modified AVAR definition \cite{Allan1981}.

An important advantage of AVAR and its more refine versions over WRMS in practical application is its weak sensitivity to low-frequency signal variations.
By contrast, WRMS depends heavily on the model used to eliminate the systematic component of the studied signal.
Our study showed that WAVAR is more robust to outliers than the classical AVAR is; however, both AVAR and WAVAR may estimate
noise level erroneously when jumps occur in a time series. 

AVAR is also widely used to investigate the spectral characteristics of a time series and is a powerful tool for noise type identification
through log-log representation.
In particular, AVAR facilitates the effective analysis of signals with different types of noise at various frequency bands.
AVAR is supposed to be more computationally effective than classical spectral methods, such as Fourier transform; in our opinion, however,
this advantage is no longer significant at present. 
A detailed comparison of two methods for estimating spectral noise characteristics may be interesting and useful though.

It must be noted that the AVAR method needs further investigation and development for some applications.
First, many geodetic and astronomical series are unevenly spaced as was discussed in Section~\ref{sect:overview}.
Frequent examples include station position time series with gaps, radio source position time series, VLBI-derived session-wise EOP series.

Another open issue regarding the application of AVAR to the analysis of geodetic and astronomical time series stems from possible correlations
between the measurements that may distort statistical analysis results substantially. 

Finally, we can conclude that despite its limitations and some unresolved issues, AVAR nevertheless remains one of the most powerful tools
for analyzing a wide range of physical measurement time series.

\section*{Acknowledgements}
This work was partly funded by the Russian Government Program of Competitive Growth of Kazan Federal University.
The author is grateful to three anonymous reviewers for careful reading of the manuscript and helpful comments.

\bibliography{my_eng,geodesy,astronomy,earth,math}
\bibliographystyle{plain}

\end{document}